\documentstyle[psfig]{l-aa}
\def\eta{et al. }

\def\ergs{${\rm erg\,cm^{-2}\,s^{-1}}$ }
\def\ergse{${\rm erg\,cm^{-2}\,s^{-1}}$}

\def\nh{$N_{\rm H}$ }

\def\src{4C\,+73.18 }

\begin{document}
\thesaurus{11.17.4:4C\,+73.18 (1928+738) -- 13.25.2}
\title{X-ray observation of the super-luminal quasar 4C\,+73.18 
(1928+738) by ASCA} 
\author{W. Yuan$^{1,2,3,4}$ \and W. Brinkmann$^{1}$ 
\and M. Gliozzi$^{1}$ \and Y. Zhao$^{3}$ \and  M. Matsuoka$^{4}$ }
\institute{
$^1$ Max--Planck--Institut f\"ur extraterrestrische Physik,
Giessenbachstrasse, D-85740 Garching, Germany \\
$^2$ Beijing Astrophysics Center (BAC),
No.1 Yifu Building 5501, Peking University, Beijing 100871 China\\
$^3$ Beijing Astronomical Observatory, National Astronomical Observatories,
        Chinese Academy of Sciences, Beijing 100012, China\\
$^4$ National Space Development Agency of Japan (NASDA), 
Tsukuba Space Center, 2-1-1 Sengen, Tsukuba, Ibaraki 305 Japan}
\date{Received Sep. 1, 1999; accepted Feb. 24, 2000}
\maketitle
\markboth{W. Yuan et al.: ASCA observation of 4C\,+73.18 (1928+738)}{}
\begin{abstract}
The results of an X-ray observation of the nearby 
core-dominated, super-luminal quasar
4C\,+73.18 (1928+738) with ASCA are reported.
Variations of the  1--10\,keV flux on time-scales of the order of 10 hours
were found while the 0.5--1\,keV flux remained almost unchanged.
The measured flux at 1\,keV had increased by $\sim 50\%$  
compared with previous observations 18 years ago. 
The quasar shows a curved spectrum in the 0.8--10\,keV band,
which can be modeled as a power law plus a soft component. 
Both, the spectrum and the uncorrelated flux variations
indicate the presence of a soft X-ray excess
over the extrapolation of the hard band power law,
which might be the high-energy extension of 
the intense UV bump emission found in this quasar.
The hard X-ray emission is dominated by the flat power law 
(photon index 1.1--1.5).
The results are consistent with models in which 
the hard X-rays are emitted from the relativistic jet. 
An iron line emission as reported 
in a previous Ginga observation (Lawson \& Turner 1997) was not detected. 
We suggest that the line detected by Ginga might be 
not associated with the quasar but, most likely, with a distant cluster
of galaxies.
There are indications for a weak emission line-like feature around 1.25\,keV 
(1.63\,keV in the quasar frame) 
in both, the SIS and GIS spectra. 
The energy is surprisingly close to  
that of an unexplained line found
in the flat-spectrum quasar PKS\,0637-752 by Yaqoob \eta (1998).
No evidence for the proposed binary black holes (Roos \eta 1993) is found in
the current X-ray data.
\end{abstract}

\section{Introduction}

4C\,+73.18 (1928+738) is an extremely core-dominated, 
super-luminal quasar at a redshift of 0.302.
It is a flat-spectrum radio source in the S\,5 survey (K\"uhr \eta 1981),
which has been well studied with VLBI and 
shows unusual jet properties. 
On arc-second (kpc) scales, the source exhibits two-sided curved jets, lobes, 
and a dominant core (Johnston \eta 1987, Hummel \eta 1992, Murphy \eta 1993). 
On pc scales, VLBI observations of the core reveal an one-sided jet,
which exhibits apparent super-luminal motion with $v_{\rm app}/c \sim 4-7$
(Eckart \eta 1985,  Witzel et al.\ 1988).
The jet is estimated to be aligned within 
$7^{\circ}-12^{\circ}$ to the line-of-sight
(e.g.\ Ghisellini \eta 1993, Jiang \eta 1998).
No $\gamma$-ray emission was detected by EGRET (Fichtel \eta 1994).
To explain the mis-alignment between the VLBI jet
and the kpc-scale jet, Hummel \eta (1992) suggested that the 
observed VLBI jet might
be only one of the Doppler-boosted filaments.
Interestingly, the most recent results from 
the space VLBI (VSOP) monitoring of 4C\,+73.18
reveal substantial jet bending at a distance of
6.3\,pc from the core, and dramatic temporal 
variations of the bend angle 
over a few months (Murphy et al.\ 1999).

It is intriguing that the sub-milliarcsec jet 
exhibits ballistic motion along a sinusoidal curve 
with a period of about 3 years (Hummel \eta\ 1992).
The wiggles over such short period were modeled 
by Roos \eta (1993) as the orbital motion of a binary black hole system,
which has a mass of the order of $10^8\,M_{\odot}$, a mass ratio $>0.1$, and 
a separation of $\sim 10^{16}$\,cm. 
No other observational evidence for the binary black hole model 
has been reported so far. 
X-ray observations, having the advantage of looking into the 
region closed to the black hole, provide a potential tool to find signatures
of binary black holes.

In the X-ray band,
4C\,+73.18 has been observed previously by {\em Einstein} 
(Biermann et al.\ 1981), 
{\em EXOSAT} (Biermann \eta 1992, Ghosh \& Soundararajaperumal 1992), 
Ginga (Lawson \& Turner 1997), 
and by ROSAT in the Survey (Voges et al.\ 1999) 
and pointed observations with PSPC
(Brunner \eta 1994, and Sambruna 1997) and HRI, respectively.
The measured spectra were generally described by a power law.
Of particular interest is the detection of 
a significant iron K$_\alpha$ emission line in the Ginga observation, 
making this object one of the few 
flat-spectrum radio quasars showing iron K$_\alpha$ line. 
Furthermore, the ROSAT data 
seem to suggest an absorption column in excess of the Galactic value. 

In this paper, we report on results of an X-ray observation
of 4C\,+73.18 in the 0.8--10\,keV band 
performed by the ASCA satellite.
The spectral and temporal analysis of the data 
are presented in \S\,2 and 3, respectively. 
A re-analysis of the archival ROSAT
PSPC spectrum is also given (\S\,\ref{spec:rosat}).
We discuss the implications of the results and summarize the main
conclusions in \S\,4 and \S\,5, respectively.  
Errors are quoted at 68\% level throughout the paper,
 unless mentioned otherwise.

\section{X-ray spectrum}
\subsection{ASCA observation and data reduction}
4C\,+73.18 was observed with ASCA (Tanaka et al.\ 1994) on 12 August, 1997.
The Solid-state Imaging Spectrometer (SIS) was operated in 1-CCD faint mode.
The data reduction and spectral analysis was performed using FTOOLS (v.4.2) 
and XSPEC (v.10), respectively.
We used the `bright~2' mode data for SIS with the corrections for 
dark frame error and echo effect applied. 
After removing hot and flickering pixels,
the data screening was performed in the standard way using the following 
criteria: an elevation angle of $10^{\circ}$, a magnetic cutoff rigidity
of 6\,GeV/c, and bright Earth angles of $25^{\circ}$ for the SIS0 and 
$20^{\circ}$ for the SIS1.
The effective exposures for good data interval were 
16.8/17.6\,ksec for the SIS0/1 and 17.8\,ksec 
for the Gas Imaging Spectrometer (GIS).
The source counts were extracted from  circular regions of $\sim 3.5$ and 6 arcmin
radius for SIS and GIS, respectively. 
The background counts were determined in two ways,
from blank sky observations at the same region of the detector 
and from a `local' off-source region of the same observation 
(at the same off-axis position with 
the same radius as the source region in the case of GIS).
The count rates of the two kinds of background data agree 
with each other within about 10\% relative errors.
In the analysis we always used both  backgrounds
and compared the results  but only those obtained by using blank sky
data are presented here, for they have a higher signal-to-noise ratio.   
The averaged net source count rates 
of the SIS1/2 and GIS3/4 are 0.14\,cts/s and 0.13\,cts/s, respectively.
The spectra of both the two GIS and SIS detectors were 
combined\footnote{We found that, for relatively weak sources, 
the spectrum combining applications
in the standard FTOOLS package resulted in unreasonably large 
statistical errors and thus unrealistically small $\chi^2$
in the spectral fits. We therefore used pure Poissonian errors
for the counts in each bin for the combined spectrum.
This might slightly overestimate the resulting $\chi^2$ of the fits
since systematic uncertainties were not taken into account;
however, for relatively weak sources, the uncertainties were dominanted
by photon statistics.}
and re-binned to have at least 30 counts in each energy bin.  
 
\subsection{The ASCA spectrum}
\label{spec:asca}
\subsubsection{A single power law fit}
\begin{figure}
\psfig{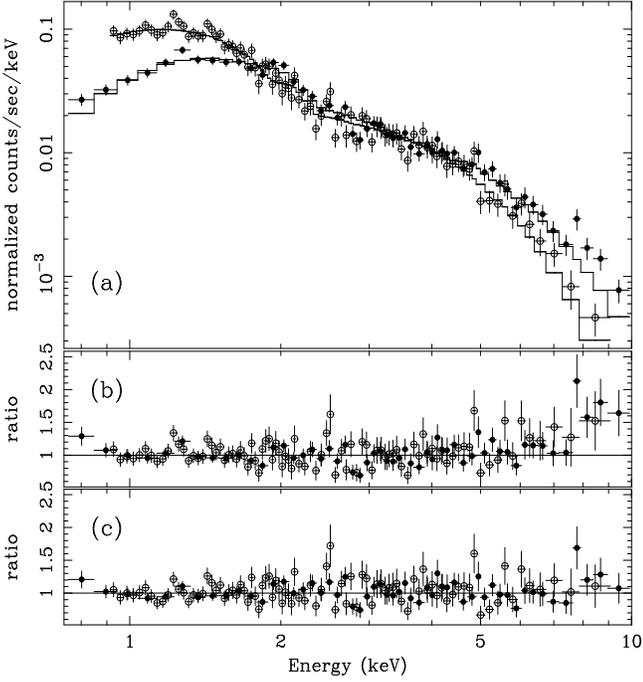}
\caption{\label{fig:one-pl} 
({\bf a}) Simultaneous fit to the SIS (open circles) and GIS (dots) spectra with
an absorbed power law (Galactic absorption). 
({\bf b}) The residuals of (a) as the ratio between 
the data and the fitted model. 
({\bf c}) The residuals of the best fit
model composed of a power law with Galactic absorption 
plus cold reflection, thermal plasma,
and an additional Gaussian line (see text).}
\end{figure}

Due to the increasing degradation of the response at low 
energies the calibration below $\sim 1$\,keV is 
uncertain for the SIS\footnote{ASCA Calibration Uncertainties, 1999, 
ASCA web page at GSFC, NASA.}.
As a conservative criterion, we ignored the energy band below
0.9\,keV\footnote{Fitting an absorbed power law to the 0.6--10\,keV SIS spectra
gave \nh significantly larger than the Galactic value ($7.4\times 10^{20}$),
$16.0\times 10^{20}$ for SIS1 and  $9.7\times 10^{20}$ for SIS0, which
are inconsistent with the GIS spectra.}
(Ikebe, private communication).
A simultaneous single power law fit to the combined SIS and GIS spectra 
is not acceptable (see Table\,1) either for free or for fixed Galactic absorption
($N_{\rm H}^{\rm Gal}=7.40 \times 10^{20}$\,cm$^{-2}$, Dickey \& Lockman 1990). 
Noticeable deviations from a power law can be seen from the residuals 
in Fig.\,\ref{fig:one-pl} (a) and (b), which show
a spectral hardening above 6\,keV and a line-like feature around 1.3\,keV.
Even ignoring the 1.1--1.5\,keV data which covers the line-like feature 
yielded no acceptable fits. 
Repeating the same fits (fixing \nh at the Galactic value) 
to the 2--10, 3--10, and 4--10\,keV band data
gave progressively flattening photon indices 
$\Gamma =1.61\pm 0.06$, $1.48^{+0.08}_{-0.10}$, and $1.37^{+0.14}_{-0.18}$, 
respectively), indicative of a curved spectral shape in the 0.8--10\,keV band
(1--13\,keV in the quasar frame).
This suggests composite spectral models, i.e.\, 
a power law plus a soft X-ray component. 

The integrated 2--10\,keV flux is 
$6.2 \times 10^{-12}$ \ergse, 
using the mean value of the SIS and GIS measurements, which 
corresponds to a K-corrected luminosity of 
$2.9\times 10^{45}$\,erg\,s$^{-1}$ ($q_{0}=0.5, H_{0}=50$).

No emission line features were seen at the iron K$_{\alpha}$ line energy.
The 90\% upper limits on the equivalent width of a 6.4\,keV line
ranges from 62 to 112\,eV (source rest frame)
for line-widths ranging from 0.01 to 0.5\,keV.

\subsubsection{Power law plus soft X-ray component}
We have tried the following models for the soft X-ray component: 
power law (PL), blackbody (BB), disk-blackbody (DBB),
and thermal plasma (RS, Raymond-Smith in XSPEC).
The results are summarized in Table\,1 except for DBB models. 
The listed temperatures  are in the quasar rest frame. 
In general, the addition of a soft component improves the fits
significantly
($\Delta \chi^2=$16, 16, 17, 19 for PL, BB, DBB, RS, respectively);  
yet the fits remain relatively poor.
Ignoring the 1.1--1.5\,keV line-like feature 
improved the fits slightly and made them  marginally acceptable, 
with a significance level 
$P\sim$ 0.11 for PL, 0.13 for BB, 0.16 for DBB, 0.25 for RS, respectively.
Thus, the current data are insufficient 
to impose strong constraints on the spectral form of
the soft X-ray component, given the degrading SIS low energy response.
On the other hand, as will be discussed below, it is also possible
that none of the above models is a good representation of 
the real spectral shape.
The thermal plasma model yields the minimal $\chi^2$ and \nh in
good agreement with the Galactic value, but,  
implausibly low metal abundances and a much flatter
photon index than the `canonical value' $\Gamma \simeq 1.6$
for flat-spectrum radio quasars. 

\begin{table}
\caption{Spectral fits for 4C\,+73.18}
\label{tab:spec_fit}
\begin{tabular}{llll} \hline\hline
parameters  & SIS+GIS       & SIS+GIS$^{(a)}$ & PSPC \\ \hline
\multicolumn{4}{l}{power law }\\
$N_{\rm H}^{(b)}$& $\sim 0$      & $\sim 0$         & 10.9$^{+1.2}_{-1.0}$  \\
$\Gamma$   & 1.69$^{+0.04}_{-0.02}$ & 1.67$^{+0.03}_{-0.02}$ & 2.25$\pm$0.12 \\
$N^{(c)}$   & 1.67$\pm$0.15 & 1.63$\pm$0.16    & 1.88$\pm$0.13 \\
$\chi^2$/dof&1.26/140       & 1.20/121         & 0.95/23       \\ \hline
\multicolumn{4}{l}{power law (fixed \nh = 7.4~$10^{20}$)}\\
$\Gamma$    & 1.79$\pm$0.02 & 1.77$\pm$0.02    & 1.82$^{+0.05}_{-0.06}$ \\
$N     $    & 1.90$\pm$0.07 & 1.84$\pm$0.09    & 1.63$\pm$0.06 \\
$\chi^2$/dof& 1.31/141      &  1.27/122        & 1.73/24 \\ \hline
\multicolumn{4}{l}{power law + power law }\\
\nh         & 26.4$\pm$15.5 & 22.8$\pm$21.5   & 13.6$^{+3.4}_{-2.6}$ \\
$\Gamma$    &1.14$\pm$0.45  & 1.12$\pm$0.60    & 1.5 (fixed) \\
$\Gamma_{\rm s}$&3.04$\pm$1.04&2.93$\pm$1.44   & 3.26$^{+0.74}_{-0.85}$\\ 
$\chi^2$/dof& 1.17/137      & 1.14/118         & 0.94/22      \\ \hline
\multicolumn{4}{l}{power law + zbbody} \\
\nh         & $\sim 0$      & $\sim 0$         & 8.9$^{+1.3}_{-0.9}$  \\
$\Gamma$    & 1.42$^{+0.05}_{-0.09}$ & 1.42$\pm$0.08    & 1.5 (fixed)   \\
$kT$ (keV)  & 0.44$^{+0.02}_{-0.04}$ & 0.44$^{+0.04}_{-0.03}$ & 0.18$^{+0.03}_{-0.02}$ \\
$\chi^2$/dof& 1.17/137      & 1.15/118         & 0.93/22       \\ \hline
\multicolumn{4}{l}{power law + zbbody (\nh = $7.4~10^{20}$)} \\
$\Gamma$    & 1.48$^{+0.03}_{-0.04}$ & 1.49$^{+0.04}_{-0.06}$ & 1.17$^{+0.18}_{-0.30}$ \\
$kT$        & 0.38$^{+0.02}_{-0.03}$ & 0.37$^{+0.03}_{-0.04}$ & 0.21$\pm$0.02 \\
$\chi^2$/dof& 1.18/138                   & 1.16/119         & 0.97/22 \\ \hline
\multicolumn{4}{l}{power law + thermal plasma } \\
\nh         & 8.1$^{+4.9}_{-2.7}$ & 12.2$^{+7.0}_{-4.2}$ & 7.3$^{+0.6}_{-0.5}$  \\
$\Gamma$    & 1.02$^{+0.22}_{-1.02}$ & 1.27$^{+0.16}_{-0.15}$ & 1.5 (fixed)  \\
$kT$        & 1.83$\pm$0.43 & 1.44$^{+0.21}_{-0.26}$  & 0.84$^{+0.18}_{-0.13}$\\
$A^{(d)}$    & 0.1$\pm$0.1  &0.5$^{+1.1}_{-0.3}$     & 1.0 (fixed)  \\
$\chi^2$/dof& 1.16/136       & 1.08/117      &   1.07/22     \\ \hline
\multicolumn{4}{l}{power law + thermal plasma (\nh = $7.4~10^{20}$)} \\
$\Gamma$    & 1.10$^{+0.13}_{-0.53}$ & 1.27$^{+0.16}_{-0.26}$ & 1.50$^{+0.17}_{-0.70}$\\
$kT$        & 1.76$^{+0.73}_{-0.21}$ & 1.59$^{+0.29}_{-0.21}$ & 0.86$^{+0.19}_{-0.15}$\\
$A$         & 0.1$\pm$0.1            & 0.4$^{+1.3}_{-0.2}$ & 0.2$\pm$0.2  \\
$\chi^2$/dof& 1.15/137       & 1.08/118      &  1.08/21   \\ \hline
\multicolumn{4}{l}{perxav + thermal plasma + Gaussian (\nh =$7.4~10^{20}$)} \\
$\Gamma$    &  1.50$^{+0.16}_{-0.50}$ & 1.41$^{+0.22}_{-0.35}$ & --  \\
$kT$        &  1.83$^{+0.40}_{-0.29}$ & 1.62$^{+0.30}_{-0.23}$ & --  \\
$A$         &  0.2$\pm$0.2            & 0.4$^{+0.8}_{-0.3}$ & --  \\    
$\Omega/2\pi^{e)}$ &0.58$^{+0.47}_{-0.58}$ & 0.46$^{+0.81}_{-0.46}$ & --  \\
$\cos \theta ^{(f)}$ &0.97$^{+0.03}_{-0.65}$ & 0.96$^{+0.04}_{-0.90}$ & --  \\  
$E_{\rm line}^{(g)}$ SIS &1.25$^{+0.03}_{-0.04}$ &  --  &  --      \\ 
$E_{\rm line}$     GIS &1.26$\pm$0.05 &  --  &  --      \\
$I_{\rm line}^{(h)}$ SIS & $2.1^{+1.1}_{-1.0}$       &     --          &     -- \\
$I_{\rm line}$     GIS & $3.7^{+2.5}_{-2.2}$       &     --          &    --  \\
$\chi^2$/dof     & 1.15/131    &  1.09/116       &    --  \\ \hline
\end{tabular}

 (a) Excluding 1.1-1.5\,keV line-like structure \\
 (b) Column density in units of 10$^{20}$\,cm$^{-2}$ \\
 (c) Normalization at 1\,keV in units of $10^{-3}$ photon\,cm$^{-2}$\,s$^{-1}$\,keV$^{-1}$\\
 (d) Abundances in units of solar value \\
 (e) Solid angle subtended by the reflecting matter\\
 (f) Inclination angle of the normal of the reflecting slab\\
 (g) line energy in units of keV\\ 
 (h) line intensity in units of $10^{-5}$\,photon\,cm$^{-2}$\,s$^{-1}$
\end{table}

\subsubsection{Reflection component}
\label{fit-refl}
An additional soft X-ray component seems not to be able to 
fully account for the spectral hardening.
In the case of thermal plasma model, 
the power law index is too flat ($\Gamma=1.10^{+0.13}_{-0.53}$),
while for the other models the residuals  show 
some remaining excess flux in the highest energy bins.
Fitting a broken power law plus thermal plasma model 
yielded a low energy band index 
$\Gamma=$1.4 and a flat index 0.1 in the hard energy range 
beyond $\sim 7$\,keV.
We then  tried to include a Compton-reflection component 
({\it pexrav} in XSPEC),
in which a Compton thick, neutral (except H and He)
slab is irradiated by an  X-ray source with 
a subtended solid angle $\Omega$.
The high-energy cutoff was fixed at 500\,keV.
The fitted photon index became $1.50^{+0.16}_{-0.50}$.
However, the improvement of the fit is insignificant and 
the addition of the reflection component cannot be justified.
It should be noted that the model of
a power law plus reflection but without a soft component
yielded no acceptable fits. 

\subsubsection{Emission line like feature around 1.3\,keV}
We now consider the emission line like feature around 1.3\,keV.
We repeated the same fitting as in \S\,\ref{fit-refl} 
(RS plus {\it pexrav} model) by adding 
a Gaussian line at $\sim 1.3$\,keV with fixed line width $\sigma=0.01$.
The line energies and normalizations were set to be independent for
the SIS and the GIS.
The best-fit parameters are given in Table\,\ref{tab:spec_fit}, 
and the residuals are shown in Fig.\,\ref{fig:one-pl} (c).
The fit improved with $\Delta\chi^2=6.0$.
Given four additional parameters involved 
(the line energies and intensities for both SIS and GIS),
the significance for adding the line component is only marginal
(at a confidence level slightly less than 90\%). 
The derived equivalent width is $22^{+12}_{-11}$\,eV for the SIS
and $31^{+21}_{-20}$\,eV for the GIS data.

\begin{figure}
\psfig{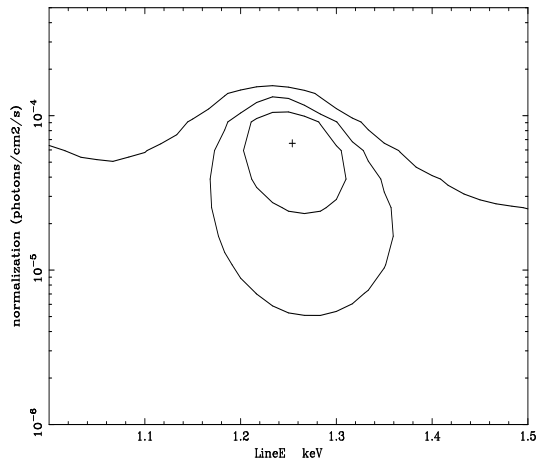}
\caption{\label{fig:line_cont}
Confidence contours in the line energy versus strength space 
for the line fit to the GIS data.
The contours are at 68\%, 90\%, and 99\% levels 
from the center outwards, respectively.}
\end{figure}

It seems unlikely that the excess in Fig.\,1 is an instrumental effect
as the feature appears in both, the SIS and GIS data at
almost identical energies,
$1.25^{+0.03}_{-0.04}$ and $1.26\pm 0.05$\,keV, respectively.
The line seems to be narrow, 
with an apparent width of the order of the detector energy resolution,
and to be non-Gaussian, as indicated by the structure of the residuals
in Fig.\,\ref{fig:one-pl} (c). 
Plotted in Fig.\,\ref{fig:line_cont} are the 
confidence contours in the parameter space of the
line energy vs.\ strength, which were obtained by using the GIS data only. 
We also tried absorption models instead of an emission line to see
whether the line-like feature can be accounted for.
However, neither an absorption edge nor a notch line fits the data.

\subsection{The ROSAT spectrum}
\label{spec:rosat}
To check the consistency of the spectral models derived from ASCA 
with previous observations, we re-analyzed the archival spectrum
of the PSPC pointed observation of 4C\,+73.18. 
A power law with neutral absorption model yields 
a steep photon index ($\Gamma= 2.25\pm 0.12$)
and absorption \nh
significantly higher than the Galactic value (Table\,1).
Fixing \nh at the Galactic value gives an unacceptable fit
($P=0.01$) and a flatter index. 
These results are in agreement 
with those given in Brunner et al.\ (1994) and Sambruna (1997).
In light of the ASCA spectral modeling,
we added a soft X-ray component and 
performed the same fits as in \S\,\ref{spec:asca}. 
The  results are given in Table\,\ref{tab:spec_fit}. 
Considering the narrow PSPC bandpass,
we first fix the photon index of the hard power law at $\Gamma=1.5$.
Adding a blackbody or thermal plasma model both yields the best fits, 
with $kT=0.18^{+0.03}_{-0.02}$ and $0.84^{+0.18}_{-0.13}$\,keV (source frame), 
respectively, and an \nh consistent with the Galactic value.
Fixing \nh at the Galactic value and setting $\Gamma$ as a
 free parameter gives  
$\Gamma=1.50^{+0.17}_{-0.70}$ for the thermal plasma and 
$\Gamma=1.17^{+0.18}_{-0.30}$ for the blackbody model, respectively. 
The fits are insensitive to the metal abundances,
and the best fit value is $0.17\pm0.16$ solar.
The excess absorption suggested previously is not required,
which apparently resulted from fitting a single power law 
to the steepening part of the curved spectrum toward low energies.

No emission line like feature is seen around 1.25\,keV in the ROSAT spectrum.
However, the poor energy resolution (40\% at 1\,keV) 
makes the PSPC insensitive to such a weak line. 
The upper limit on the equivalent width 
of a narrow emission line at 1.25\,keV is 
found to be 74.5\,eV at 90\% confidence,
which is consistent with the ASCA data.

\subsection{Summary of spectral fits}

The ASCA spectra exhibit a concave curvature which 
flattens toward high energies.
Models composed of a power law and a soft X-ray excess 
improve the fits significantly,
though the spectral shape of the soft component is inconclusive. 
Such models also fit the previous ROSAT spectrum well, 
hence the soft X-ray excess in 4C\,+73.18 seems to be 
persistent, rather than a transient emission feature.
No significant excess X-ray absorption is indicated.
No iron K$_{\alpha}$ line  is detected. 
A weak emission line like feature at 1.25\,keV
is present in both, the SIS and GIS spectra,
though the significance is not high.
It should be noted that the ASCA spectra are `noisy' above 2\,keV,  
and the spectral fits are relatively poor in general.  
This may result, at least partly, from  
source variability as shown below. 

\section{X-ray variability}
\subsection{Variability on short time-scales} 
\begin{figure}
\psfig{file=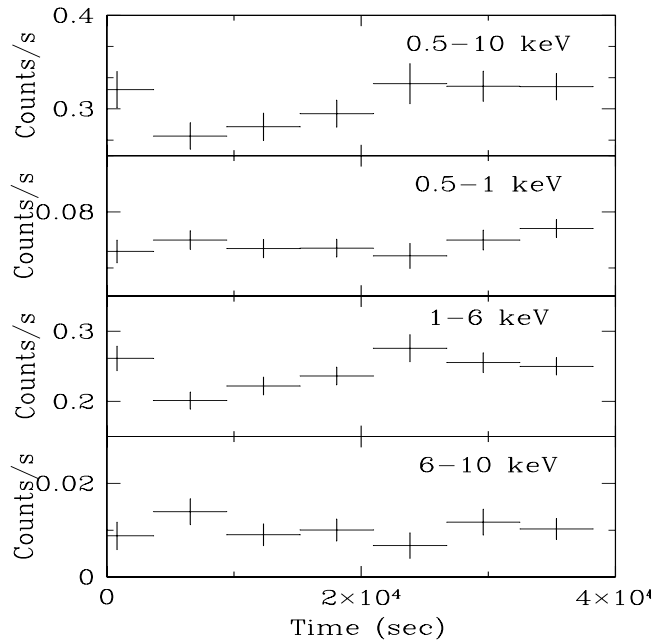,width=8.5cm,height=9.0cm,angle=0}
\caption{\label{fig:vari_flux} Background subtracted X-ray light curves of \src 
from  the SIS detectors (SIS0+SIS1) in the 
whole band (upper panel) as well as
in the low, medium, and high passbands
The bin size is 5760\,s and the errors are of 1\,$\sigma$.}
\end{figure}

X-ray light curves were extracted from the source region and binned
with a bin size of 96\,min (one ASCA orbit) for both the SIS and GIS detectors.
The background count rates, estimated from the source free region and 
normalized to the source region, were subtracted.
Plotted in Fig\,\ref{fig:vari_flux} are the light curves 
from the SIS detectors only. 
Small amplitude, but statistically significant flux variations in 
the 0.5--10\,keV band are 
found ($\chi^2$ test, see Table\,\ref{tab:vari}) 
over the $\sim40$\,ksec observation.
As shown above, the X-rays in the ASCA band 
are likely composed of a hard power law, a soft component dominating
below 1\,keV, and a tentative reflection component 
above 7\,keV (9\,keV in the source frame).
We divided the ASCA energy band into three corresponding bands, i.e.\  
a low (0.5--1\,keV), a medium (1--6\,keV), and a high (6--10\,keV) band.
The light curves in Fig.\,\ref{fig:vari_flux} show different temporal
variations in the medium band (the power law component) 
from those in the low and high bands. 
The medium band flux reveals significantly more variability than 
the whole band ($\Delta \chi^2=4$), 
whereas the low and high energy band fluxes are consistent with 
no variations (Table\,\ref{tab:vari}). 
We checked the significance of the 
relative variations between different bands
by testing the constancy of their counts ratios,
$C_{\rm 1-6keV}/C_{\rm 0.5-1keV}$ and  $C_{\rm 1-6keV}/C_{\rm 6-10keV}$.
The probabilities for the $\chi^2$ tests 
are listed in Table\,\ref{tab:vari}.
Non-synchronized variations between the medium and 
low, and the medium and high bands, however, cannot be confirmed. 
This is likely due to the poor photon statistics in the low and high bands,
as well as to the contribution from the underlying power law flux
in these two bands.
The GIS light curves show very similar variations;
however, the large counting errors due to the higher background level
makes the  GIS insensitive to variability with  such small amplitudes.
\begin{table}
\caption{$\chi^2$ test of variability for the SIS fluxes}
\label{tab:vari}
\begin{tabular}{llll} \hline\hline
band (keV) & $\chi^2$ (dof=6) & $P (\chi^2)$ & significance \\ \hline
0.5--10  & 13.5 & 0.035 & yes \\
0.5--1   & 5.9  & 0.43  & no \\
1--6     & 17.4 & 0.008 & yes \\
6-10     & 4.3  & 0.64  & no \\ \hline
\multicolumn{4}{l}{Variability of counts ratio} \\ \hline
$C_{\rm 1-6keV}/C_{\rm 0.5-1keV}$ & 8.1 & 0.23 & no \\
$C_{\rm 1-6keV}/C_{\rm 6-10keV}$  & 5.3 & 0.50 & no \\ \hline
\end{tabular}
Note: we use the critical significance level of $P=0.05$.
\end{table}

We conclude that the observed short time scale flux variability in \src
is mainly attributable to variations of the hard band power law.
The variability time scale is not obtainable from  
the relatively short
observation interval; however, it could be of the order of the 
duration of the observation, i.e.\ 10 hours in the source frame. 
No statistically significant variations are found for the flux in 
the 0.1--1\,keV band.
This result is supported by the lack of variations of
the flux in the 0.1--2.4\,keV PSPC band
on similar time scales during the ROSAT observation
($\chi^2=1.06$ for 37 d.o.f., Brunner et al.\ 1994).
The uncorrelated, energy dependent flux variations are consistent with the
above models of multi-component X-ray emission.

\subsection{Long-term variability} 

\begin{figure}
\psfig{file=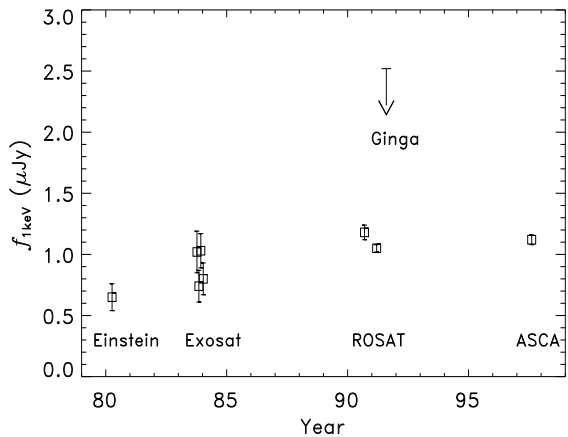,width=8.5cm,height=6.5cm,angle=0}
\caption{\label{fig:hist_lc} Historical light curve of the X-ray 
flux densities of 4C\,+73.18 measured at 1\,keV.
The fluxes of {\em Einstein} and {\em EXOSAT} 
observations are taken from Brunner et al.\
(1994). The two ROSAT data points represent the Survey and pointed
observations, respectively.
The estimated Ginga flux (taken as an upper limit here; see text) 
is an extrapolation of the measured 
2--10\,keV spectrum from Lawson \& Turner (1997).}
\end{figure}

We present in Fig.\,\ref{fig:hist_lc} 
the long-term X-ray light curve of 4C\,+73.18, compiled from  
the flux densities at 1\,keV from previous missions.
The estimated Ginga flux is an extrapolation of
the observed 2--10\,keV spectrum 
and, as discussed below, should be taken as 
an upper limit due to possible 
flux contamination from nearby sources. 
The light curve tends to suggest an overall increase of 
the 1\,keV flux density 
by a factor of 50\% over 18 years of observations.

In addition, prominent flux variations in the hard X-ray band
were reported from the {\em EXOSAT} observations:
the 2--10\,keV flux varied by a factor of 2--3 
(from 1.5--5.1$\times 10^{-12}$\,\ergs) 
on a time scale of about one month
(Ghosh \& Soundararajaperumal 1992).

\section{Discussion}

\subsection{Emission from host cluster of galaxies?}
As mentioned by Brunner \eta (1994, quoting a private communication) 
there are indications for a 
cluster of galaxies around 4C\,+73.18,  which would affect 
the X-ray spectrum determined from instruments with limited spatial resolution.
To investigate the possible role played by an  
extended X-ray emission component, we fitted a point source plus
a $ \beta$-model (e.g. Gorenstein et al. 1978) of the form
$$
S(r)=S_0\left(1+{r^2\over r_c^2}\right)^{-3\beta+1/2}
$$
to the surface brightness profile of the ROSAT HRI source 
(with an exposure of 38660\,sec).
The result is inconclusive; however, a  noticeable contribution
from a cluster at a flux level greater than a few percent
in the HRI energy band can be ruled out. 

In the hard X-ray band (0.8--10\,keV), 
no evidence for extended emission 
is found in the ASCA GIS image.
Moreover, no iron K$_{\alpha}$ emission line is detected in the ASCA spectra.
It should be noted that the GIS source extraction radius of 6\,arcmin 
corresponds to $\sim 1.3$\,Mpc at the quasar distance, 
which is about 2--3 times the typical X-ray core radius
and hence may well encompass the X-ray emitting region of the host
cluster.
Taking the typical value of 1\,keV for 
the equivalent width of the iron K$_{\alpha}$ line of cluster emission, 
the derived $EW<50$\,eV places an upper limit on the flux
density of a host cluster of being 7\% of the quasar flux at about 5\,keV
in the observer frame.
Thus, we do not find evidence for 
noticeable X-ray emission from a host cluster of galaxies.

\subsection{The iron line puzzle}
\label{dis:line}
One of the interesting results of this work is the non-detection of
the iron K$_{\alpha}$ line, which was previously reported 
from the Ginga observation (Lawson \& Turner 1997).
The upper limits of the equivalent width is estimated to be 
62\,eV, in contrast to the measured
$185^{+79}_{-74}$\,eV by Ginga.
Moreover, the spectrum ($\Gamma=2.08^{+0.05}_{-0.04}$)
is steeper and the flux 
($8.7\times10^{-12}$\,\ergse) higher
for Ginga than for the ASCA  observations 
($\Gamma=1.61\pm0.06$ and flux $6.2\times10^{-12}$\,\ergse)
in the 2--10\,keV band.
The difference of the two continuum spectra can be seen clearly in 
Fig.\,\ref{fig:sed}.
There are two possible explanations to this observed discrepancy, namely,
either genuine spectral variations of both the iron line and continuum, 
or contamination of the Ginga spectrum by other sources.
We consider the latter to be more plausible,  
based on the following arguments.

\subsubsection{Iron line variability}
It is rare for blazar-type radio quasars to show iron line emission 
(e.g.\ Siebert et al.\ 1996, Lawson \& Turner 1997, 
Reeves et al.\ 1997, Cappi et al.\ 1997).
X-ray monitoring of 3C\,273 also revealed that the iron line
was detected in only some but not all of the observations.
This effect seems to be dependent on variations of the
continuum rather than of the line flux---at a high continuum flux,
the line equivalent width decreases or the line is even swamped
(e.g.\ Turner et al.\ 1990, Cappi et al.\ 1998).
However, this was not the case for 4C\,+73.18 during the Ginga observation
(see below for a discussion on the Ginga flux).
If the line did come from the quasar,
there must be a variation of the line itself.
This may suggest that during the Ginga observation Seyfert like 
emission from the central AGN was substantially enhanced 
while the jet emission dimmed since 
no significant changes were found for the overall flux level.

\subsubsection{Iron line from another object?}
Given the large field of view (FOV) of $1.1^{\circ}\times2.0^{\circ}$ (FWHM)
and poor energy resolution of Ginga, 
the association of the detected line with the quasar, and even 
the presence of the line, may be questionable.
We examined X-ray sources in the vicinity of 4C\,+73.18 which 
may have fallen into the FOV of the Ginga observation.
Within the GIS FOV there is another source detected at a distance of
12\,arcmin from the quasar;
however, the flux in the Ginga energy band is too low
($2.7\times10^{-13}$\,\ergse) to
make considerable contribution to the observed Ginga flux.

We next analyzed the data of ROSAT PSPC pointed observation of 4C\,+73.18,
whose $2^{\circ}$ FOV covers a large portion of the Ginga FOV.
Among 27 X-ray sources detected (with a detection likelihood larger than 10) 
in addition to 4C\,+73.18, we found one source of particular interest.
The source is located 
43\,arcmin away from the quasar and 
inside the Ginga FWHM field of view (Yamagisi, private communication).
It is bright (0.1 counts\,s$^{-1}$ in the 0.1--2.4keV band---40\% of 
the count rate of the quasar) 
and extended (4 times of the FWHM of the point spread function).
This source has been selected from the ROSAT All-Sky Survey as
a candidate of, and later confirmed in optical to be, a cluster of galaxies
(Boehringer et al.\ 2000).
The ROSAT spectrum, which is of poor photon statistics, 
can be fitted with thermal bremsstrahlung emission,
and the temperature seems to be higher than 3\,keV. 
Though the redshift is unknown, we suggest 
this X-ray cluster as a likely source of the emission line detected by Ginga.

As a self-consistency check, we compare the 
line intensity predicted for the cluster with the observed value.
The observed line flux by Ginga is 
$2.55^{+1.08}_{-1.02}\times10^{-5}$\,photon\,s$^{-1}$\,cm$^{-2}$ 
and the line energy
$5.13^{+0.30}_{-0.20}$\,keV (observer frame).
Assuming a cluster temperature of 6\,keV and 
extrapolating the fitted ROSAT spectrum to higher energies, 
we expect that Ginga would detect a line flux of
$F_{\rm line}\simeq 4.4 f \, EW/(1+z) \times 10^{-5}$\,photon\,s$^{-1}$\,cm$^{-2}$,
where $EW$ is the line equivalent width 
(in units of keV, in the source frame)
and $f$ the Ginga off-center detection efficiency at the source position 
($f\simeq 0.6$ in this case).
For a redshift close to that of the quasar, $z\sim 0.3$, we have  
$F_{\rm line}\sim 2.0 EW$ ($10^{-5}$\,photon\,s$^{-1}$\,cm$^{-2}$),
which is compatible with the observed value for a range of line 
equivalent width from several hundred eV to 1\,keV. 
In this case, the measured Ginga line energy  
implies that the cluster is at a redshift range 0.23 $ \leq z \leq $ 0.36
for a 6.7\,keV K$_{\alpha}$ line.
However, no firm conclusion can be drawn on the origin of the line 
until the redshift of the cluster is available.

\subsubsection{The Ginga continuum flux}
The estimated continuum flux from the cluster, using the above parameters,
is $\sim 2\times 10^{-12}$\,\ergs in 2--10\,keV, 
which may account for the measured flux excess for Ginga over the
ASCA measurement. Further,
though the rest of the sources in the PSPC FOV are too faint and soft
to be considerable contributors to the observed Ginga flux, 
they may account for the steepening of the Ginga spectrum in the soft
X-rays.
The total PSPC count rate of sources in the vicinity of 
4C\,+73.18 turned out to be $\sim 0.36$\,counts~ s$^{-1}$,
1.3 times higher than that of quasar itself.
Thus we suggest that the excess fluxes of the Ginga observation
over those of ROSAT and ASCA in the soft and medium 
X-ray band (see Fig.\,\ref{fig:sed})
could be, at least partly, accounted for 
by emission from the cluster of galaxies and other 
nearby soft X-ray sources.  
The genuine X-ray flux of the quasar might be comparable with that
of ASCA and ROSAT, though quantitative estimates can only be obtained 
by detailed modeling. 

\subsection{The soft X-ray excess}
\begin{figure}
\psfig{file=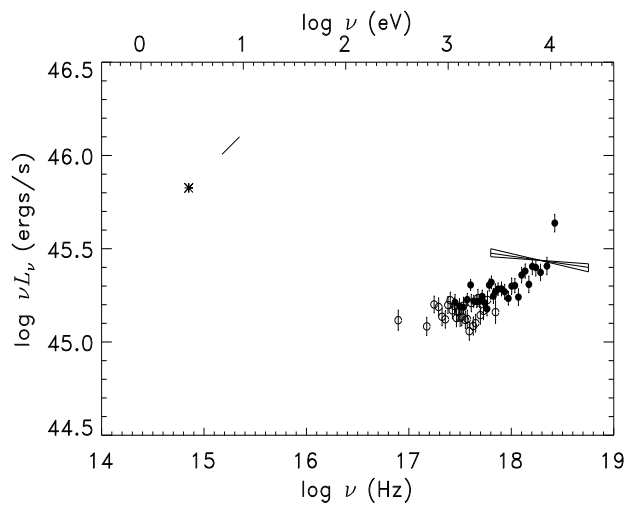,width=9.0cm,height=7.0cm,angle=0}
\caption{\label{fig:sed} The optical to X-ray spectral energy
distribution of 4C\,+73.18 in the source rest frame ($q_{0}=0, H_{0}=50$).
The UV ({\it HST}) spectrum is taken from Laor et al.\ (1995). 
The open circles are the ROSAT data and the filled dots the ASCA GIS data. 
For a comparison, the power law spectrum 
obtained in the Ginga observation is over-plotted. 
}
\end{figure}

The presence of an excess flux at energies below $\sim 1$\,keV 
over the extrapolation of the hard band spectrum is evident
from the curved spectrum in the ASCA band
and the uncorrelated variations in the soft (0.5--1\,keV) and 
hard (1--6\,keV) band. 
The latter was also reported from the {\em EXOSAT} observations
(Ghosh \& Soundararajaperumal 1992).
However, the spectral shape of the soft excess is uncertain. 
We show in Fig.\,\ref{fig:sed} 
the optical to X-ray energy distribution of 4C\,+73.18, which 
is very similar to that of 3C\,273. 
The optical-UV spectral shape, though with sparse data points, 
is suggestive of 
a prominent UV component which peaks at shortward 1300\,$\AA$
or above $10^{15.3}$\,Hz (the `blue bump'),
and probably extends into the EUV band.
It is natural to regard  the soft X-ray excess 
to be the high-energy tail of the UV bump,
which results from Compton up-scattering disk photons 
to higher energies by hot gas.

When fitted with a thermal emission model,
the temperature $kT_{\rm ex}$ (e.g.\ 0.4\,keV for blackbody)
is somewhat higher than those obtained in a few 
core-dominated radio quasars
for which $kT_{\rm ex}$ was relatively well determined,
e.g.\  $\sim 0.1$\,keV in 3C\,273 (e.g.\ Turner \eta 1990,
B\"uhler \eta 1994, Cappi \eta 1998).
The higher $kT_{\rm ex}$ could imply that the up-shift of 
the energy of disk photons is more pronounced 
in 4C\,+73.18 than in 3C\,273.
The hot gas may be the skin-type coronae surrounding the accretion disks,
as discussed in detail in, e.g., Czerny \& Elvis (1987).
In the Czerny \& Elvis model, 
the high $kT_{\rm ex}$ implies that 
the quasar is likely radiating at a luminosity 
around the Eddington luminosity
(see Fig.\,7 of Czerny \& Elvis 1987). 
If this is true, 
a UV luminosity of the order $\sim 10^{46}$\,ergs\,s$^{-1}$ 
(see Fig.\,\ref{fig:sed}) implies a central mass of 
the order of $8\times 10^7$\,M$_{\odot}$ in 4C\,+73.18.
This value is close to the mass of the binary black hole system 
($10^8$\,M$_{\odot}$)
estimated by Roos et al.\ (1993) by modeling the wiggles of the VLBI jet.
Other sources of the hot gas might be
the base of the relativistic jet where it emerges from the 
accretion disk, as proposed by Mannheim \eta (1995).
These scenarios all require a small inclination angle of the disk
(or the jet, respectively) 
to the observer, in agreement with the super-luminal motion 
observed in this source.
Finally, the result is consistent with the findings by Brinkmann et al.\ (1997) 
that the soft X-ray excess tends to be more significant 
in core-dominated than in lobe-dominated quasars. 

In the soft X-ray band, the spectral analyses in \S\,2 indicate a
higher temperature for the
soft thermal component in the ASCA than in the ROSAT observations.
This may well be a genuine spectral change,
as the observations are about 6 years apart.
The conclusion, however, must be drawn with caution due to
the inter-instrument calibration uncertainties.
There have been reports that the ROSAT PSPC tends to give steeper spectra
than ASCA, even within the overlapping bandpass (e.g.\ Iwasawa et al.\ 1999).
Such an effect might play a role
in comparing the temperatures derived from the two instruments.

\subsection{The hard X-ray continuum: emission from jet}

The observed flat power law continuum in the hard X-ray band 
is typical for flat-spectrum radio-loud quasars. 
The inferred time scale of variations of the power law continuum
might be as short as of the order of hours.
These factors, as well as the lack of an iron $K_{\alpha}$ line, 
are typical signatures of
the radio-jet-linked X-ray emission in radio-loud AGN, 
which is strongly beamed by relativistic bulk motion. 
Thus the ASCA observation provides strong evidence for 
the idea that the hard X-rays in 4C\,+73.18 are emitted predominantly 
from the jet (via inverse Compton scattering).
Such mechanism has been tested in detail in a few similar super-luminal
quasars (e.g.\ 3C\,345, Unwin \eta 1994).

To investigate the contribution of possible Seyfert-like emission,
we tried to include a steep power law (fixing $\Gamma=1.9$)
as well as reflection in the above fittings.
No acceptable fits were obtained when a significant
Seyfert-like emission component was included.
Though, as shown in \S\,\ref{spec:asca}, 
including a reflection model can account for the tentative
excess fluxes in the highest energy bins of the ASCA spectra,
the addition of this component is not justified statistically by the data.
Given the lack of an iron line and that the continuum is
dominated by beamed jet emission, a significant reflection component
can not be physically justified either.

\subsection{The 1.3\,keV emission line feature}

A marginal, narrow line feature at 1.25\,keV  
is suggested to be present in 4C\,+73.18. 
It is not clear from the current data 
whether this is a real line or an instrumental artifact,
though the latter seems unlikely given that this feature appears 
in both the co-added SIS and GIS data at almost identical energies.
If interpreted as associated with the quasar, the line energy corresponds to
1.63$\pm0.05$ and 1.64$\pm0.07$\,keV 
in the quasar frame for the SIS and GIS, respectively.  
No known emission lines have been found or are 
expected to be present around these energies.
The closest line is the He-like MgXI transition at 1.58\,keV.

Interestingly enough, using ASCA data Yaqoob \eta (1998) 
reported the discovery of a 
peculiar narrow emission line ($EW=59^{+38}_{-34}$\,eV),
which has never been seen in any other AGN, 
in another flat-spectrum quasar PKS\,0637-752 (redshift 0.654).
The line energy in quasar frame is at 1.60$\pm0.07$, which 
is strikingly close to what we found in this work.
Therefore, the line-like feature in 4C\,+73.18 provided
the first independent evidence for the presence of such an emission line,
though the statistical significance of the presence of the line is low.
It seems unlikely that the match of the line energies 
in two quasars with different redshifts is a coincidence. 
The identification of this line is, however, unknown 
(see Yaqoob \eta 1998 for a discussion of the possibilities).
If the emission line feature found in this work is real and identical to that
of Yaqoob \eta (1998), 
the obtained $E_{\rm line}\sim 1.6$\,keV 
should be the line energy in the rest frame,
rather than a Doppler-shifted value due to the in- or out-flow of the 
emitting gas.
Otherwise it is hard to explain why the bulk motions
are  exactly the same in the two quasars.

\section{Conclusions}
The flat-spectrum, super-luminal quasar 4C\,+73.18 
resembles the well studied, similar object 3C\,273 
in some basic properties in the soft-to-medium X-ray band. 
Rapid, small amplitude ($\sim 25\%$) flux variations 
were detected in the 1--10\,keV band on a tentative time scale
of the order of 10 hours,
whereas no significant variations were found for 
the soft X-rays below 1\,keV. 
The 0.8--10\,keV spectrum exhibits a curvature,
which can be modeled by a power law plus a soft component.
Both, the X-ray spectrum and the uncorrelated flux variations
provide strong evidence for the presence of a soft X-ray excess 
over the extrapolation of the hard band power law,
which might be the high-energy extension of the UV bump emission. 
The underlying continuum is flat, 
with a power law photon index of 1.1--1.5,
depending on the spectral model adopted for the soft component.
The dominance of the flat power law 
over possible Seyfert-like emission, 
together with the short time scale variability,
are naturally explained by beaming models for the jet,
which is estimated to be observed at an inclination angle about 10 degree. 
No iron K$_{\alpha}$ emission line is detected, in contrast to 
the significant line reported in a previous Ginga observation,
which is most likely contamination from another source.
A weak emission line-like feature around 1.25\,keV
(1.6\,keV in the quasar frame)
seems to be present in both the SIS and GIS spectra, 
though the significance is not high.
The line energy is identical to that of a mysterious emission line
in the flat-spectrum quasar PKS\,0637-752 reported by Yaqoob \eta (1998).

\begin{acknowledgements}
WY thank Y.\ Ikebe for the help regarding the ASCA data analysis and
discussion on the X-ray cluster in the Ginga FOV, 
and also thank S.\ Xue for useful discussions.
The authors thank I.\ Yamagisi at ISAS for getting the attitude information 
of the Ginga observation.
We thank the Northern ROSAT All-Sky (NORAS) team for providing the
identification results on the cluster of galaxies before publication.
WY is grateful to Prof.\ J.\ Tr\"umper and Dr.\ W.\ Voges for the 
financial support during the visit at MPE, and also to 
Beijing Astronomical Observatory for the financial support and
to Prof.\ J.\ Chen at Beijing Astrophysics Center for hospitality,
where part of the research was done.
WY acknowledges support at NASDA by a STA fellowship.
MG  acknowledges partial support from
the European Commission under contract number ERBFMRX-CT98-0195
(TMR network ``Accretion onto black holes, compact stars and
protostars").

\end{acknowledgements}


\begin{thebibliography}{}
\bibitem{} Biermann P.L., Duerbeck H., Eckart A., et al., 1981, ApJL 247, 53 
\bibitem{} Biermann P.L., Schaaf R., Pietsch W., et al., 1992, A\&AS 96, 339
\bibitem{} Boehringer H., et al.\ 2000, ApJ submitted
\bibitem{} Brinkmann W., Yuan W., Siebert J., 1997, A\&A 319, 413
\bibitem{} Brunner H., Lamer G., Staubert R., Worrall D.M., 1994, A\&A 287, 436
\bibitem{} B\"uhler P., Courvoisier T.J.-L., Staubert R., 1994, A\&A 287, 433
\bibitem{} Cappi M., Matsuoka M., Comastri A., et al., 1997, ApJ 478, 492
\bibitem{} Cappi M., Matsuoka M., Otani C., Leighly K.M., 1998, PASJ 50, 213
\bibitem{} Czerny B., Elvis M., 1987, ApJ 321, 305
\bibitem{} Dickey J.M., \& Lockman F.J. 1990, ARA\&A 28, 215
\bibitem{} Eckart A., Witzel A., Biermann P., et al., 1985, ApJL 296, L23
\bibitem{} Fichtel C.E., Bertsch T.L., Chiang J., 1994, ApJS 94, 551
\bibitem{} Ghisellini G., Padovani P., Celotti A., Maraschi L., 1993, ApJ 407, 65
\bibitem{} Ghosh K.K., Soundararajaperumal S., 1992, MNRAS 254, 563
\bibitem{} Gorenstein P.D., Fabrikant D., Topka K., Harnden F.R., Tucker W.H., 1978, ApJ 224, 718
\bibitem{} Hummel C.A., Schalinski C.J., Krichbaum T.P., et al., 1992, A\&A 257, 489
\bibitem{} Iwasawa K., Fabian A.C., Nandra K., 1999, MNRAS 307, 611
\bibitem{} Jiang D.R., Cao X.W., Hong X., 1998, ApJ 494, 139
\bibitem{} Johnston K.J., Simon R.S., Eckart A., et al., 1987, ApJL 313, 85
\bibitem{} K\"uhr H., Pauliny-Toth I.I.K., Witzel A., Schmidt J., 1981, AJ 86, 854
\bibitem{} Laor A., Bahcall J.N., Jannuzi B.T., Schneider D.P., Green R.F., 1995, ApJS  99, 1 
\bibitem{} Lawson A.J., Turner M.J.L., 1997, MNRAS 288, 920
\bibitem{} Mannheim K, Schulte M., Rachen J., 1995, A\&A 303, L41
\bibitem{} Murphy D.W., Browne I.W.A., Perley R.A., 1993, MNRAS 264, 298
\bibitem{} Murphy D.W., Tingay S.J., Preston R.A., et al., 1999, Elsevier preprint
\bibitem{} Reeves J.N., Turner M.J.L., Ohashi T., Kii T., 1997, MNRAS 292, 468
\bibitem{} Roos N., Kaastra J.S., Hummel C.A., 1993, ApJ 409, 130
\bibitem{} Sambruna R., 1997, ApJ 487, 536
\bibitem{} Siebert J., Matsuoka M., Brinkmann W., Cappi M., et al., 1996, A\&A 307, 8
\bibitem{} Tanaka Y., Inoue H., Holt S.S., 1994, PASJ 46, L37
\bibitem{} Turner M.J.L., Williams O.R., Courvoisier T.J.-L, 1990, MNRAS 244, 310
\bibitem{} Unwin S.C., Wehrle A.E., Urry C.M., et al., 1994, ApJ 432, 103
\bibitem{} Voges W., Aschenbach B., Boller Th., et al., 1999, A\&A 349, 389
\bibitem{} Witzel A., Schalinski C.J., Johnston K.J., et al., 1988, A\&A 206, 245
\bibitem{} Yaqoob T., George I.M., Turner T.J., et al., 1998, ApJL 505, 87
\end{thebibliography}
\end{document}